\documentclass[aps,prl,address,twocolumn]{revtex4}
\usepackage{amsmath}
\usepackage{amssymb}
\usepackage{graphicx}

\newcommand{\ignore}[1]{}               

\begin{document}
\title{Ticking terahertz wave generation  in attoseconds}
\author{Dongwen Zhang, Zhihui L\"{u}, Chao Meng, Xiyu Du, Zhaoyan Zhou, Zengxiu Zhao}\email{zhao.zengxiu@gmail.com}
\author{Jianmin Yuan}\email{jmyuan@nudt.edu.cn}
\address{Department of Physics,
National University of Defense Technology, Changsha 410073, P. R. China}
\begin{abstract}
We perform a joint measurement  of terahertz waves and high-order harmonics  generated from noble atoms driven by a fundamental laser pulse and its second harmonic. 
By correlating their dependence on the phase-delay of the two pulses,  we  determine the generation of  THz waves in tens of attoseconds precision. 
Compared with simulations and models, we find that   the laser-assisted soft-collision of the electron wave packet with the atomic core  plays a key role. It is demonstrated that the rescattering process, being indispensable in HHG processes,  dominant THz wave generation as well but in a more elaborate way.  The new finding might be helpful for the full characterization of the rescattering dynamics.
\end{abstract}
\date{\today}
\maketitle

\vspace{8mm}

Intense terahertz (THz) radiation from ionizing gases in two-color laser fields has attracted great interest recently not only for the application such as remote sensing \cite{Liu10} but also for understanding the mechanism of terahertz generation \cite{Cook00, Kress04, Kress06, Xie06, Kim07, Kim08NP, Silaev09, Wen09, ChenYanping09, Dai09,Batushkin10}. When the fundamental and frequency-doubled femtosecond laser pulses are focused in gases, the intensity of the THz radiation is modulated by the relative phase between the two pulses. Experiments confirm that THz output increases dramatically with the onset of  plasma formation, indicating that ionization plays a key role in THz generation \cite{Kress04, Kress06}. This led to involve four-wave mixing (FWM) in plasma to elucidate the extreme nonlinear phenomenon but without the knowledge of the origin of   $\chi^{(3)}$ responsible for such high efficiency. 

  A transient photoelectron current model is recently proposed taking into account of the electron dynamics following ionization \cite{Kim07, Kim08NP}.
 In this model,  electrons are released from atoms through tunnel ionization and then driven by non-sinusoidal two-color laser field which breaks  the dynamic symmetry of the electronic wave-packet. A directional current is thus formed causing the emission of THz waves.  However, there is a discrepancy between the above models on the relative phase when the terahertz yield maximizes. 
It is evident that the key problem of describing THz generation in two-color fields is to find the buildup process and the subsequent relaxation of the transient current \cite{ Kim08NP, Batushkin10}. The sensitivity  on the phase delay indicates that the sub-cycle electron dynamics  needs to be measured  to understand THz generation and  a determination  of the relative phase in attosecond precision is urgently expected.

 It is known that a commensurate two-color field can be used to steer trajectories of continuum electrons and control their re-scattering/re-collision with the parent ion to manipulate  high-order harmonic generation (HHG)  \cite{Dudovich06N, Mashiko08, Feng09}. In HHG processes, each harmonic can be uniquely related to two quantum orbits within each optical cycle  based on the rescattering theory \cite{Lewenstein94}.  Two-photon ionization experiments have confirmed that the emission time of harmonics differs in attoseconds  and it varies linearly with harmonic order in positive slope for the short trajectories \cite{Paul01, Mairesse03, Aseyev03}.  The  weak harmonic field tunes the phase difference of  radiation  in the positive and negative half cycle of the fundamental, causing the appearance of even-order harmonics with intensity modulated by  phase delay of the two pulses. The two-color scheme thus  provides an all-optical in situ method for  measuring the birth of XUV pulses \cite{Dudovich06N} which is an ideal tool  for ticking the generation of THz in attoseconds. Furthermore,  since both yields of THz waves and harmonics modulate when the time-delay of the two pulses varies by  attoseconds,  a question arises:  are they related knowing that THz waves have duration of picoseconds?

We therefore perform  a joint measurement  by monitoring THz  and high-order harmonic  yields in a two-color laser field simultaneously.  
By correlating their phase-delay dependence,  the absolute relative phase dependence of THz yields can be retrieved and it is found that  
THz yields  take maximum  at phase delay of  $0.8\pi$ deviated from the prediction of previous models. Compared with numerical  simulations and theoretical models, we find that the Coulomb potential from the residue ion should be taken into account and  the  laser-assisted soft-collision of the electron wave packet with the atomic core  affects  the phase dependence of Thz yields.   Thus  terahertz radiation provides an complementary  tool to study the electron dynamics in addition to high-order harmonic generation and above threshold ionization, helping the full characterization of  the rescattering wave packet. 
 
In the experiment, we used a 1-kHz Ti:Sapphire chirped pulse amplification (CPA) system that delivered 1.6 mJ at 790 nm in a 25 fs pulse. 
The beam is reduced to a diameter of 10 mm and split into pump and probe beams.  The pump is focused on a 200 $\mu$m thick $\beta$-barium borate (BBO) crystal (type I) 
to  generate the second harmonic beam with 20\% conversion efficiency.  A two-color Mach-Zehnder interferometer is employed to control the relative phase between the two pulses, with a pair of fused-silica wedges  in the fundamental path.  Neutral density filters are used in both arms to control their intensities separately.
High-order harmonics and THz waves are generated in the 1.5 mm long gas cell in 12 mbar by the two-color pulse using an off-axis parabolic Au-coated mirror with focal length of 200 mm. The gas cell is placed 2.5 mm behind the focus to optimize the HHG yields.
The THz waves are  reflected with a hole-drilled off-axis parabolic mirror and then combined with the delayed probe beam into a 1mm thick (110) oriented ZnTe crystal for electro-optic sampling to the waveform  below 3 THz. 
The harmonics passing the hole  is  simultaneously recorded by a home-made flat-field spectrometer for each phase-delay.

\begin{figure}
\includegraphics*[width=8cm,clip=true]{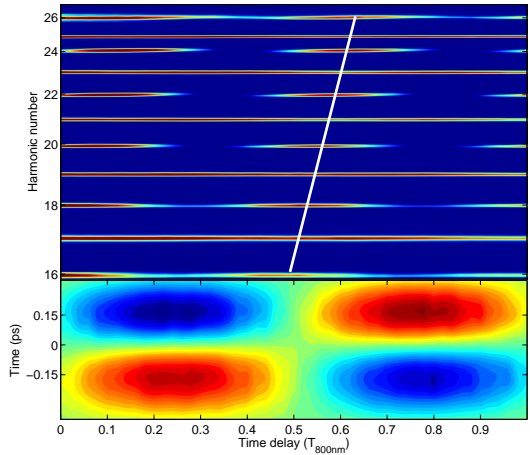}
 \caption{ Contours of the terahertz wave forms (bottom) and the intensities of harmonics  (top) from 16th to 26th varying with the time  delay of the two-color laser field, in unit of the fundamental optical period. The intensity ratio of the second-harmonic to the fundamental field is about $0.5\%$. }
\label{THz_HHG}
\end{figure}

\begin{figure}
\includegraphics*[width=6cm,clip=true]{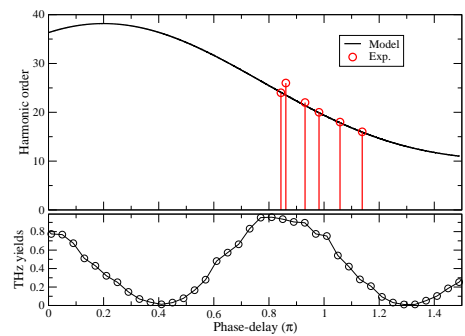}
 \caption{Top: The phase-delay maximizing intensity of even-order harmonics extracted from the experiment (circles) and theoretical model (solid line).  Bottom: modulation of THz yields with phase-delay.   The laser intensity is estimated  of $2.3$$\times$$10^{14}$W/cm${^2}$.}
\label{Corr}
\end{figure}

In Fig.~\ref{THz_HHG} we show the measured THz wave form and the yields of high-order harmonics  modulated with the time delay of the two-color laser field which is characterized by $E(t)=E_1\cos\omega t+E_2\cos(2\omega t+\phi)$, where the field strengths of the fundamental and the second-harmonic are denoted by $E_1$ and $E_2$ respectively, and the phase delay is related to the time delay by $\phi=2\omega t_d$.   The  solid line depicts the phase-delay $\phi_0(N\omega)$  maximizing the yields of  even-order harmonics from 16th to 26th.   In order to determine the absolute phase-delay,  we apply the procedure given in \cite{Dudovich06N} by mapping $\phi_0(N\omega)$ to the emission time of harmonics obtained from the rescattering model \cite{Lewenstein94}.  
The intensity of 2N$^{th}$ order harmonic  is modulated approximately  as $ \sin^2{\Delta}$, where $\Delta$ is  the additional action introduced by the vector potential of the second-harmonic pulse and is given by \cite{Dudovich06N}
\begin{equation}
\Delta \approx   \frac{\lambda E_1^2}{2\omega^2}\int_{t_i}^{t_r} (\sin\omega t-\sin\omega t_i)\sin(2\omega t+\phi)dt
\end{equation}
with the ionization instant $t_i$ and the recombination instant $t_r$ as  functions of harmonic energy and $\lambda$ is the ratio of $E_2$ to $E_1$.
The corresponding phase-delay that maximizes 2N$^{th}$ harmonic yields is plotted in solid lines  in Fig.~\ref{Corr}. The laser intensity is estimated  of  $2.3$$\times$$10^{14}$W/cm${^2}$ from the cut-off energy of harmonics. Also shown in Fig.~\ref{Corr} is the extracted experimental data  denoted in circles. The lower order harmonic is taking maximum at later delay in accordance to the harmonic generation from short trajectories \cite{Paul01}.  Note that the left part of the theoretical result (solid line) corresponds to the case of harmonic generation from the long trajectories which has been removed by phase-matching in experiments.  The overall agreement between theory and experiment for all the harmonics presented allows us to determine the absolute phase-delay maximizing yields of harmonics, e.g.,  the 22rd  harmonic takes maximum at $0.93\pi$. 
Note less than $0.04\pi$ phase-delay control is achieved in  the experiment corresponding to 60 attoseconds.

Now using the harmonic modulation as ticks we can assign the absolute phase-delay to the modulation of THz yields (summed yields below 3 THz) shown in Fig.~\ref{Corr}. Because the THz and harmonic yields are measured jointly for each given phase-delay, we can in fact lock the modulation of THz yield to one of the harmonics.  The THz yields are found taking maximum at phase delay of $0.8\pi$. Using the quantum mechanical approach \cite{ ZhouZY09, Silaev09},  we obtain the same result which deviates from the predication of four-wave-mixing and directional photoelectron current models. What is the reason behind?

According to the photoelectron current model, the micro current ($\vec{j}(t_i)$) generated from  the ionization event at time $t_i$ can be obtained by weighting  the final outgoing electron velocity with  the corresponding instantaneous ionization rate \cite{Kim07}. The initial velocity of the electron born at time $t_i$ is assumed zero and the final electron velocity can be found from the classical motion to the electron in laser fields. The total current responsible for THz emission is thus the coherent summation over all the ionization events.
Because the micro current generated from  two ionization events separated by $\pi/\omega$ has symmetry of $\vec{j}(t_i)=-\vec{j}(t_i+\pi/\omega)$,  their destructive interference  causes vanishing total currents for THz generation in a single color laser field. When a second harmonic pulse is introduced, the symmetry between $t_i$ and $t_i+\pi/\omega$ is broken, directional current is  formed giving rise to the emission of THz waves.

Depending on the instants  it is set free, the electron  can either directly escape from the atom or recollide with the atomic core. We can therefore define the escaping currents (EC)  and rescattering currents (RC), $\vec{j}_{esc}(t_i)$ and $\vec{j}_{res}(t_i)$ accordingly.
Because of the large energy gained from the laser field by electrons approaching the core, HHG is mainly contributed by the high frequency components of the RC. On the contrary, THz wave has photon energy in meVs,  it can be emitted at much large distances from the nucleus involving free-free transitions \cite{ZhouZY09}. Therefore the contributions from low energy and less accelerated electrons can not be ruled out  and the EC needs to be considered as well. 
Note that  EC/RC is related to tunnel ionization occurring in the quarter cycles before/after the electric field maxima.  Because  $\vec{j}_{res}(t_i)=-\vec{j}_{esc}(-t_i)$, the magnitude of the total current from a two-color laser pulse takes maximum at the relative phase of $\pi/2$ \cite{Kim08NP}.

In the above discussion regarding free electron motion, only the electron-light interaction is considered and the influence  of the Coulomb potential from the residue atomic core is neglected completely. This  so-called strong field approximation \cite{Lewenstein94} has been widely used in modeling processes such as HHG where high kinetic energy electrons are involved.
However for near-threshold phenomena, it can distort the free-electron wave function significantly and modify the density of states \cite{Perry88}. Its focusing effects  enhance double ionization by one order of magnitude \cite{Brabec96}.  The Coulomb-laser coupling has been shown affecting laser-assisted photoionization \cite{Smirnova07B}, ellipticity-dependence of low order harmonics \cite{Ivanov96}, angular distribution   \cite{Kaminski96} and low-energy peaks of of above-threshold ionization (ATI) photoelectrons \cite{Quan09} as well.
Since THz wave generation involves of slow electron motion which can be strongly distorted by the Coulomb potential, the response of the whole wave packet in the combination of the laser field and the Coulomb potential needs to be investigated. In the following we show that the Coulomb potential has different impacts on the formation of the EC and RC. 

\begin{figure}
\includegraphics*[width=8cm,clip=true]{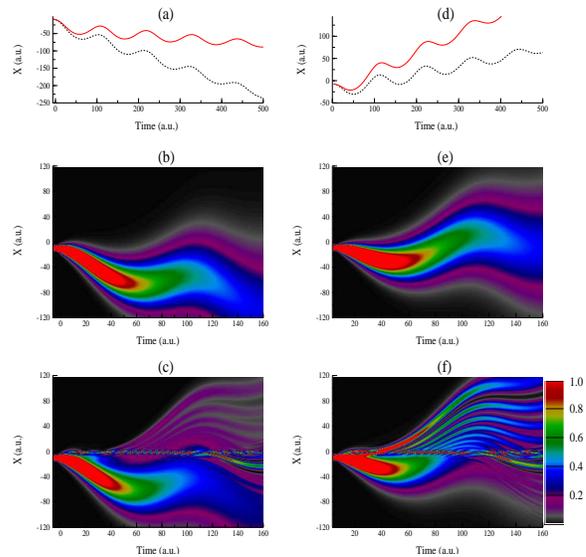}
 \caption{Time propagation of electron wave packets after tunneling ionization at two instants $t_1$ (left column) and $t_2$ (right column) corresponding to the cases of escaping and rescattering respectively. 
 The contours plots show 
the normalized  densities of probability  as functions of time and coordinates,  encoded with color scale 1.0 corresponding to the density of 0.02 a.u..  In (b) and (e), the Coulomb potential is neglected, but being considered  in (c) and (f). 
 In (a) and (d),  the dotted lines represent the classical trajectories of electrons  with zero initial velocity, and  the solid lines represent the quantum mechanical averaged electron displacements. 
   The intensities of the fundamental pulse and its second harmonic are $1\times10^{14}$ W/cm$^2$ and $5\times10^{12}$ W/cm$^2$ respectively, with zero phase delay.  Both pulses have durations of 25 fs with Gaussian envelop centered at zero time. }
\label{ResWP}
\end{figure}

We consider two ionization instants $t_1$ and $t_2$ in  the two-color laser pulse with zero phase-delay.
The  corresponding classical  trajectories of the electron driven  by  the laser field only are   shown in the dotted lines in Fig.~\ref{ResWP} (a) and (d) respectively. Quantum mechanically, we propagate the continuum electron wave packet (EWP) following tunneling ionization.
The initial electron wave packets at the instants of tunneling ionization are assumed to have the same shape of the ground state wave function but shifted to the classical turning points  at  $x_0=-I_p/E_f$, where $I_p$ is the ionization potential and $E_f$ is the corresponding instantaneous electric field.  One-dimensional time-dependent Schr\"{o}dinger equation is solved for simplicity. 
The evolving  of the EWP's  in the absence of Coulomb potential  are shown in figure.~\ref{ResWP}(b) and (e)  respectively.  As one of the quantum features, the EWP's are broadened during evolution because of  the quantum dispersion. 
However, the center motion of the  EWP coincides exactly with the classical motion of the electron   in a laser field with zero initial momentum,
thanks to the conservation of the canonical momentum in both the classical and quantum theories. 


It can be seen in Fig.~\ref{ResWP}(e) that the wave packet moves toward the negative $X$ axis initially and then turns back to the positive axis  passing the origin  where the atomic core is positioned.   It is clearly the scenario of rescattering. The central part of the wave packet makes hard collision with the atomic core  that  causes high-oder harmonic generation and  high energy above-threshold ionization when followed by either recombination or back-scattering \cite{Paulus94}.  
However, the  THz yields are determined by the ejecting photoelectron current, therefore we must consider the propagation of the wavepackt after scattering by the atomic potential. We see in Fig.~\ref{ResWP} (f) the electron wave packet is fragmented after re-collided with the atomic core.  Especially the upper wing of the electron wave packet in Fig.~\ref{ResWP} (f) is strongly enhanced in density comparable to that of the center of the electron wave packet.    The magnitude of the averaged velocity is increased because of more fractions of electron moving toward the positive direction. The averaged displacement  plotted in solid line in Fig.~\ref{ResWP}(d) demonstrates the photoelectron current  is indeed enhanced. Different from recombination and backscattering processes,  here we are considering  the evolution of the whole wave packet with most of the packet scattered  away from the atom. In other words, it is the laser-driven soft collision of the EWPs with atomic core causing the enhancing of the rescattering current. 
It might be  the same mechanism that causes low-energy peaks  in photoelectron spectra \cite{Kastner12}.  

 On the contrary in Fig.~\ref{ResWP}(b), the center of the EWP  moves away from the origin in a manner of quiver motion, without the chance to reencounter the atomic core making hard collisions.  Due to the broadening of the EWP, fractions of electrons in the wing of the EWP pass through the origin, 
 but the averaged velocity of the wave packet makes no change when the atomic potential is ignored. If the Coulomb potential is included in the time propagation we see in Fig.~\ref{ResWP}(c) that the wing of the EWP is altered by the atomic core.  The EWP is fragmented as well with more fractions of electrons moving  toward  the positive axis  by the attractive Coulomb potential.  Because the  total drift velocity is negative in  the absence of Coulomb potential, it is suppressed as plotted in the solid line compared with the dotted line in Fig.~\ref{ResWP} (a). Here we see again that the soft-collision with the atomic core has profound effect on the EC. Note  EC makes no contribution in the theory of HHG \cite{Lewenstein94}.  

 The laser-assisted soft collision is mainly occurring within the first quarter cycle following ionization because the electron drift away from the nucleus at later time such that the effect of Coulomb potential is not significant any longer. The momentum transfer is contributed most from the time  that the electron makes the turn where its velocity is zero such that it has more time interacting with the  Coulomb potential. It can be investigated theoretically from the classical electron motion in the combination of the atomic potential and the laser field \cite{Smirnova07B, Smirnova08}.  In the scenario  of rescattering, our estimation gives approximate momentum transfer of $\Delta p\approx2E\cos\theta/\kappa^3$, where $\kappa=\sqrt{2I_p}$  with $I_p$ of the atomic ionization potential and $E\cos\theta$ is the field at instant of ionization. 
 Because of the momenta transfer by the Coulomb potential, the symmetry between the EC and the RC is broken with  the former  suppressed and the RC enhanced. Therefore  the rescattering process  responding for HHG, dominates  THz wave generation as well but because of the laser-driven soft-collision of EWP with atomic core.
 
\begin{figure}
\includegraphics*[width=8cm,clip=true]{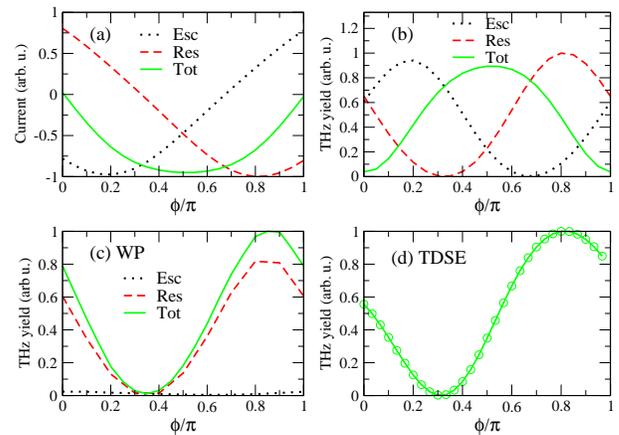}
 \caption{ (a) Escaping, rescattering  and total currents calculated from the photocurrent model and (b)  the corresponding THz yield below 3 THz; (c) THz yields obtained from classical wavepacket simulation by considering Coulomb potential; (d) THz yields calculated from solving the time-dependent Schr\"{o}dinger equation with the same laser parameters  given in Fig.~\ref{ResWP}.}
\label{current}
\end{figure}

 In order to further check the analysis above, we perform classical and quantum calculation for the THz yields from a two-color laser pulse. As shown in Fig.~\ref{current}(a) and (b), the THz yield takes maximum at phase-delay of $0.5\pi$ from the photocurrent model which ignores the Coulomb potential. When the Coulomb potential is taken into account in the classical wave packet simulation shown in Fig.~\ref{current}(c), the THz yield from the EC is suppressed  and that from the RC is enhanced. 
 In particularly, the total THz yield takes maximum at the phase-delay of $0.8\pi$ agreed with one dimensional quantum mechanical calculation shown in  Fig.~\ref{current} (d) confirming the previous discussion. In the classical wave packet simulation, we sample and trace the trajectories starting with Gaussian distributed transversal momenta  according to the tunneling ionization theory \cite{Ammosov86}.

Finally, the phase-delay dependence of the RC can be obtained analytically by treating the second harmonic as a perturbation. It yields (details will be given elsewhere)
 \begin{equation}
 J\approx\frac{e^2}{m}\frac{E_2}{\omega}e^{-\alpha}\big\lbrace  \frac{\alpha-1}{\alpha}\cos\phi-\sqrt{\frac{9\pi}{8\alpha}}\sin\phi   \big\rbrace,
 \end{equation}
 where $e$ and $m$ represent the charge and the mass of the electron respectively, and $\alpha={2\kappa^{3}}/{3E_1}$, where  ionization rate  $w=\frac{4}{E_1}e^{-\alpha}$for hydrogen-like atom  is used \cite{Ammosov86}.
The generated THz wave field strength  is oscillating with $\phi$ at the period of $2\pi$ which corresponding to time delay of ${\pi}/{\omega}$. It can be seen that  THz wave takes maximum when  $\tan\phi=\sqrt{\frac{9\pi}{8}}\frac{\sqrt{\alpha}}{1-\alpha}$. This critical phase is mainly determined by  the ionization potential and the intensity of the fundamental laser pulse.  For weak $E_1$ when $\alpha \gg 1$,  the maximum yield occurs at zero phase agreed with the perturbative FWR   theory. 
For the intensity of the fundamental we uses, the  maximum occurs at phase of $0.8\pi$ agrees with our measurements.
For intensity much higher the phase approaches  to $\pi/2$ in accordance  to the transient current model \cite{Kim08NP}. 
 Therefore the two mechanisms are  channelled  depending on the competition of the soft-collision  and  the electron-light scattering. 

In conclusion, the joint measurement of THz and harmonic yields allows us  to determine the generation of THz waves in attosecond precision. While the time scales of the two kinds of radiation are quite different, we show that in fact they are connected and both  originate  from the same electron dynamics: laser-driven electron motion in the atomic fields.
Particularly we found that THz generation is sensitive to the Coulomb potential of the residue ion after tunneling ionization. The magnitude of current from directly escaping electrons is suppressed while that from the rescattering electrons is enhanced due to the soft-collision of electron wave packet with the ion.  As a first step toward fully understanding the correlation of THz generation and HHG, we show that  rescattering electron dynamics might bridge the two radiation processes.    While HHG reflects the motion of the center of electron wave packt which makes hard collision with the atomic core, THz generation reflects the whole dynamics of the electron wave packet. Therefore the THz wave provides a tool to investigate the whole electron wavepackt dynamics, beyond the recombination in HHG and back-scattering in above threshold ionization.
We expect that the sensitivity of THz yields on Coulomb potential could be used to map atomic fields from within  and help the full characterization of  the rescattering wave packet. 

This work is supported by  the National NSF of China under Grants No. 11104352, the Major Research
plan of National NSF of China (Grant No. 91121017), the National High-Tech ICF Committee of China.

\end{document}